\documentclass[pre,twocolumn,superscriptaddress,preprintnumbers,amsmath,amssymb]{revtex4-1}
\usepackage{epsfig}
\usepackage{amsmath}
\usepackage{amsfonts}
\usepackage{graphicx}
\usepackage{fancybox}
\usepackage{subfigure}
\usepackage{color}
\usepackage[titletoc]{appendix}
\usepackage{sidecap}

\newcommand{\degree}{^{\circ} }
\newcommand{\erf}{\ensuremath{\mathop{\rm erf}}}

\newcommand{\rb}{\mathbf{r}}

\newcommand{\avg}[1]{\left<#1\right>}
\newcommand{\len}[1]{\left|#1\right|}
\newcommand{\brac}[1]{\left[#1\right]}
\newcommand{\curly}[1]{\left\{#1\right\}}
\newcommand{\para}[1]{\left(#1\right)}

\newcommand{\esw}{\ensuremath{\varepsilon_{sw}}}
\newcommand{\RHS}{\ensuremath{R_{\rm HS}}}
\newcommand{\RHStilde}{\ensuremath{\tilde{R}_{\rm HS}}}
\newcommand{\tmu}{\ensuremath{\theta_{\mu}}}
\newcommand{\toh}{\ensuremath{\theta_{\rm OH}}}

\newcommand{\ie}{\emph{i.e.}}

\newcommand{\rhoq}{\ensuremath{\rho^q}}
\newcommand{\rhoqr}{\ensuremath{\rho^q_{\rm R}}}
\newcommand{\rhor}{\ensuremath{\rho_{\rm R}}}

\newcommand{\rhob}{\rho_{\rm B}}

\newcommand{\phiRLJ}{\ensuremath{\phi_{\rm R}^{\rm LJ}}}
\newcommand{\phiRlLJ}{\ensuremath{\phi_{\rm R1}^{\rm LJ}}}

\newcommand{\Vr}{\ensuremath{\mathcal{V}_{\rm R}}}

\newcommand{\phiRltilde}{\ensuremath{\tilde{\phi}_{\rm R1}}}

\newcommand{\kT}{\ensuremath{k_{\rm B}T}}

\newcommand{\E}{\ensuremath{\mathcal{E}}}

\begin{document}

%\preprint{}

\title{Dissecting Hydrophobic Hydration and Association}


\author{Richard C. Remsing}
\affiliation{Institute for Physical Science and Technology and
Chemical Physics Program, 
University of Maryland, College Park, MD 20742}
\email[]{rremsing@umd.edu}
  \author{John D. Weeks}
  \affiliation{ Department of Chemistry and Biochemistry,
  Institute for Physical Science and Technology, and
Chemical Physics Program, University of
  Maryland, College Park, MD 20742}
  \email[]{jdw@umd.edu}
%\thanks{}


%\date{\today}

\begin{abstract}
We use appropriately defined short ranged reference models of liquid water 
to clarify the different roles local hydrogen bonding,
van der Waals attractions, and long ranged electrostatic interactions 
play in the solvation and association of apolar solutes in water.
While local hydrogen bonding interactions dominate hydrophobic effects
involving small solutes, longer ranged electrostatic and dispersion interactions
are found to be increasingly important in the description of interfacial structure
around large solutes.
The hydrogen bond network sets
the solute length scale at which a crossover in solvation behavior between
these small and large length scale regimes is observed. 
Unbalanced long ranged
forces acting on interfacial
water molecules are also important in hydrophobic association, illustrated here by
analysis of the association of model methane and buckminsterfullerene solutes.
\end{abstract}
\maketitle

\section{Introduction}

Hydrophobic interactions play a key role in phenomena ranging from
biological processes like protein folding and membrane formation
to the design of water-repellent materials~\cite{BallReview,ChandlerNatureReview,DewettingRev}. 
Thus, significant effort has been
devoted to studying the behavior of apolar moieties in water. In pioneering work, Stillinger argued that hard sphere solutes smaller than a critical radius $R_C$  can be inserted into liquid water while maintaining
the hydrogen bond network, but for solutes with a radius larger than $R_C$ bonds must be broken,
generating a molecular scale interface with properties resembling that of the liquid-vapor interface in
water~\cite{StillingerLV}. More recent work has confirmed the basic features of this idea and put the arguments on a firmer statistical mechanical foundation~\cite{ChandlerNatureReview, DewettingRev,LCW,WeeksAnnRevPhysChem}.

While this qualitative description of the length scale dependence of hydrophobic hydration
seems physically very reasonable, it focuses only on the hydrogen bond network of water and makes
no mention of the van der Waals (VDW) attractions and long ranged multipolar interactions
between water molecules or of the VDW attractions that would be present between a more realistic solute and the solvent.
Moreover, a qualitatively similar length scale transition is seen in a dense Lennard-Jones (LJ)
fluid near the triple point, with the formation of a ``dry'' vapor-like interface around a large hard sphere
solute~\cite{WeeksAnnRevPhysChem}. In that case clearly there are no hydrogen bonds and the transition
is generated solely by unbalanced VDW attractive forces
arising from solvent molecules far from the solute.

Consideration of such unbalanced forces is an essential ingredient
in the Lum-Chandler-Weeks (LCW) theory of hydrophobicity~\cite{LCW}, which uses the same basic
framework to describe hard sphere solvation in simple liquids and in water, 
differing only in the thermodynamic parameters needed as input to the theory~\cite{LCW,HuangChandlerPRE,scalingHphobFreeEnergies}. Indeed LCW theory has been criticized for not
treating hydrogen bonds and other distinctive features of water more explicitly and there has also been considerable debate about possible effects of
solute-solvent LJ attractions on the proposed length scale transition in water~\cite{DewettingRev}.
Thus it seems useful to explore in more detail the varying roles hydrogen bonds, VDW interactions, and long ranged
multipolar interactions play in hydrophobic solvation, and to determine what analogies exist to
solvation in simple, non-associating fluids.

To that end, we build on our previous work~\cite{JStatPhys} using truncated water models~\cite{Nezbeda},
and exploit the underlying ideas
of perturbation~\cite{WCA,WidomScience} and local molecular
field~\cite{WeeksAnnRevPhysChem, LMFDeriv} (LMF) theories of uniform and nonuniform
fluids, respectively, to study hydrophobic solvation
and association from small to large length scales. We employ short ranged
variants of the SPC/E water model to show that small scale solvation and association
in water is governed by the energetics of the hydrogen bond network alone.
However when the solute is large
and the hydrogen bond network is broken at the hydrophobic interface,
water behaves in a manner qualitatively similar to a simple fluid, with unbalanced
LJ attractions dominating the solvation behavior.

In the next section,
the truncated water models are briefly introduced
and our simulation methods are detailed.
Section III examines the roles of unbalanced dispersion and
electrostatic forces in determining the equilibrium solvation structure around
small and large apolar solutes. The strength of the hydrogen bond network
around small solutes is then analyzed by perturbing the hydration shell
in Section IV. The role of this network in setting the length-scale for
the crossover in solvation thermodynamics is then studied
in Section V. The origin of entropy convergence is briefly discussed in Section VI.
Finally, the hydrophobic association of model methane
and fullerene molecules is studied in Section VII. Our conclusions and
a discussion of the implications of this work are given in Section VIII.

\section{Models and Simulation Details}

Hydrogen bonds in most classical water models arise from ``frustrated charge pairing'', where an effective positive charge on a hydrogen site of one molecule tries to get close to a negatively charged acceptor site on a neighboring molecule~\cite{JStatPhys}. This strong attractive interaction is opposed by the overlap of the repulsive Lennard-Jones (LJ) cores and the presence of other hydrogen sites in the acceptor molecule.
As a result, short ranged versions of the full water model where Coulomb interactions are truncated at distances larger than the hydrogen bond length and with only truncated LJ core interactions if desired can still give a very accurate description of the hydrogen bond network and pair correlation functions in bulk water~\cite{JStatPhys,LMFWater}.

In this work, we use the extended simple point
charge (SPC/E) model of water~\cite{SPCE} and two
previously developed short ranged
variants of this model~\cite{JStatPhys} to examine hydrophobic
hydration and association as the solute perturbs the hydrogen bond network. The truncated models provide a hierarchical framework for disentangling
in such classical models 
the separate contributions of (i) strong short ranged interactions leading to the hydrogen bond network, (ii) longer-ranged VDW attractions between water molecules and with the solute, and (iii) long ranged dipolar interactions between water molecules.

The first such model, the \textit{Gaussian-truncated}
(GT) water model, has full LJ interactions
but truncated Coulomb interactions~\cite{JStatPhys,LMFWater}. It
thus lacks the long ranged electrostatics
necessary to provide a description of the physical multipolar interactions
that act over large distances. We also utilize the \textit{Gaussian-truncated repulsive-core} (GTRC)
model~\cite{JStatPhys}, where both the long ranged electrostatic
interactions and the long ranged LJ attractions
have been removed. The GTRC model generates a minimal reference network model that captures very well the
structure of the local hydrogen bond network in bulk water while ignoring effects
of the remaining long ranged Coulomb and dispersion interactions.

In order to compare the SPC/E water model at a pressure of $P=1$~atm 
with the short ranged GT and GTRC models in the work presented below,
the latter two models were simulated at corrected pressures yielding the same density using
the pressure corrections described earlier~\cite{JStatPhys, TruncCoul}.
There it was
shown that simple analytical corrections to the pressure can bring
the bulk densities of these three models into quantitative agreement.
All data presented in this work were obtained from
molecular dynamics simulations performed in the isothermal-isobaric
ensemble (constant NPT) using a modified
version of the DL\_POLY2.18 software package~\cite{dlpoly}. Constant
temperature and pressure conditions were maintained through the
use of a Berendsen thermostat and barostat,
respectively~\cite{BerendsenBaroThermo}.
The evaluation of electrostatic interactions in simulations of
the full SPC/E water model employed the Ewald summation
method~\cite{CompSimLiqs}.  

It is instructive to compare the solvation behavior of water to that of
a simple LJ fluid
at an analogous state point throughout this work. Therefore,
following the work of Huang and Chandler~\cite{HuangChandlerPRE}, 
we also study a LJ fluid at a state point near the triple point, where the
potential is truncated and shifted at $2.5\sigma$. This LJ fluid is studied
at a reduced temperature and pressure of $T^*=k_BT/\epsilon=0.85$ and 
$P^*=P\sigma^3/\epsilon=0.022$, respectively, corresponding to a reduced density
of $\rho^*=\rho\sigma^3=0.70$. In order to study the analogous short ranged reference fluid, 
we use the same repulsive force truncation of the LJ potential as
was used for the GTRC water model, and study the model at a mean-field corrected pressure
that accounts for the lack of LJ attractions
~\cite{JStatPhys}.

We should emphasize that the above-mentioned short ranged GT and GTRC models are \textit{not}
being used in this paper as replacements for standard long-ranged models such as SPC/E or to give accurate 
representations of most properties of real water. Rather, we utilize these models as \textit{analysis tools}
to examine the different roles the hydrogen bond network as described by the GT or GTRC models, long ranged dispersions, and dipolar interactions play
in determining the properties of systems containing liquid water. 

However, the GT model  describes very well pair correlation functions and hydrogen bond statistics in bulk water,
and as we discuss further below, it also captures many features of the water density in
nonuniform environments including the basic length scale transition for hydrophobic solutes~\cite{JStatPhys}.
But thermodynamic and  particularly electrostatic properties depend sensitively on the long ranged Coulomb
interactions and GT results need corrections for quantitative accuracy.
Acharya and Garde~\cite{Acharya} have recently carried out a detailed study of the strengths
and weaknesses of the GT model as a simple water model in a variety of settings, including both hydrophobic and ionic solvation.

\section{The Influence of Long Ranged Interactions on Interfacial Structure}

In this section, we examine the role of the various unbalanced forces 
in determining the interfacial structure of water near a hydrophobic solute.
The solute is considered to be a uniform density of
LJ particles, such that its interaction with water can be represented by an integration of the 
LJ potential over the volume of the solute, resulting in the integrated ``$9-3$''
potential of Huang and Chandler~\cite{HuangChandlerSoluteSolvAttr}
\begin{eqnarray}
&U_{\rm sw}&\para{r;R_S}=\pi\esw\rho\sigma_{\rm sw}^3\nonumber \\
&\times&\Bigg[\frac{4}{5}\sigma_{\rm sw}^9\para{\frac{1}{8rr_+^8}-\frac{1}{9r_+^9}-\frac{1}{8rr_-^8}+\frac{1}{9r_-^9}} \nonumber \\
&-&2\sigma_{\rm sw}^3\para{\frac{1}{2rr_+^2}-\frac{1}{3r_+^3}-\frac{1}{2rr_-^2}+\frac{1}{3r_-^3}}\Bigg],
\end{eqnarray}
where $r_\pm=r\pm R_S$. The parameters of the potential are chosen to mimic paraffin,
such that the density of LJ sites, energy,
and length scale are given by $\rho=0.0240$~\AA$^{-3}$, $\esw=0.882$~kJ/mol, and
$\sigma_{\rm sw}=3.468$~\AA, respectively~\cite{HuangChandlerSoluteSolvAttr}. 
Furthermore, in order to make this particle as hydrophobic as possible,
only the repulsive part of the potential is used, such that the
solute-water interaction potential used in the MD
simulations is given by
\begin{eqnarray}
U_{0,\rm sw}(r)= \Bigg\{ \begin{array}{ll}
U_{\rm sw}(r)-U_{\rm sw}\para{r_0}, \  r\le r_0 \\ 
0, \ \ \ \ \ \ \ \ \ \ \ \ \ \ \ \ \ \ \ \ \ \ \  r > r_0\\ \end{array}
\end{eqnarray}
where $r_0$ is the location of the minimum of the potential. Finally, we should note that the size of the 
particles is better represented through an effective hard-sphere radius, $\RHS$, rather than the size parameter
$R_S$ found in the potential. This effective radius can be estimated as~\cite{Blip}
\begin{equation}
\RHS\approx\int_0^\infty dr \curly{1-\exp\brac{-\beta U_{0,\rm sw}(r)}},
\end{equation}
where $\beta=\para{\kT}^{-1}$, and will be reported as $\RHS$ herein.

%@@@@@@@@@@@@@@@@@@@@@@@@@@@@@@@@@@@@@@@@@@@@@@
\begin{figure}[tb]
\centering
\includegraphics{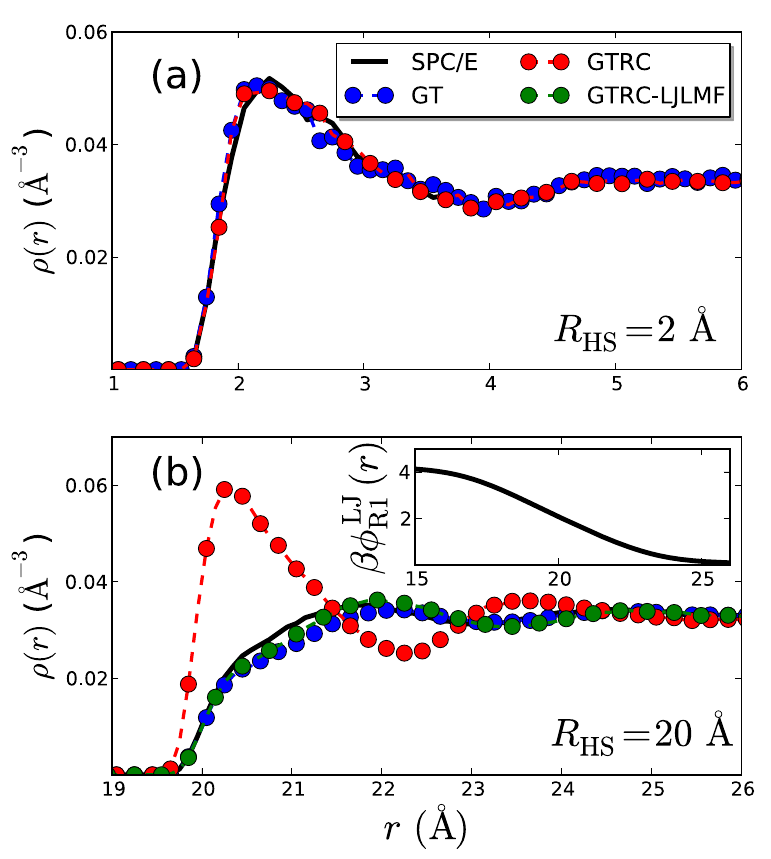} 
\caption{Density distributions around solutes of radii $\RHS\approx2$~\AA \ (a) 
and $\RHS\approx20$~\AA \ (b). The inset in (b) depicts the renormalized
portion of the LJ LMF for GTRC water in units of $\kT$.}
\label{fig:densfig}
\end{figure}
%@@@@@@@@@@@@@@@@@@@@@@@@@@@@@@@@@@@@@@@@@@@@@@

The hydration structure around small solutes has been 
postulated to be a direct consequence of the need for water to maintain its hydrogen bond
network. A small solute can be ``inserted'' into bulk water with the network continuing around the solute without breaking hydrogen bonds.
Indeed, in the small solute regime we find that the nonuniform densities of GT and GTRC models around
an apolar particle are nearly identical to that of the full SPC/E water, dramatically
confirming that local hydrogen bonding dictates the hydration 
structure in this limit (Figure~\ref{fig:densfig}a).

In the large solute limit, Figure~\ref{fig:densfig}b, the density profiles of SPC/E and GT water are still very similar,
demonstrating that long ranged electrostatic interactions
have an almost negligible influence on this measure of interfacial structure. GTRC water, on the other hand,
has a $\rho(r)$ markedly different from that of SPC/E water.

Removal of the LJ attractions from the bulk liquid in GTRC water eliminates the phenomena of drying,
and it evidentially wets the surface of the solute. 
According to LMF theory \cite{LMFDeriv}, we can account for the averaged effects of the neglected LJ forces by using a renormalized solute field
\begin{equation}
\phiRLJ(\rb)=U_{0,\rm sw}(r)+\int d\rb' \brac{\rhor(\rb')-\rhob}u_1\para{\len{\rb-\rb'}},
\label{eq:ljlmf}
\end{equation}
where quantities obtained in the presence of the effective field
are indicated by the subscript `R' throughout this work, $\rhob$ is the bulk density of the fluid,
and $u_1(r)$ is the attractive portion of the LJ potential.
The use of this renormalized field recovers drying behavior and
brings the density profile of GTRC water into qualitative
agreement with that of the SPC/E and GT models, 
as illustrated by the curve labeled `GTRC-LJLMF' in Figure~\ref{fig:densfig}b.
The renormalized portion of the LMF,
$\phiRlLJ(\rb)\equiv\phiRLJ(\rb)-U_{0,\rm sw}(r)$, provides an effective force that pushes solvent molecules
away from the solute, as shown in the inset of Figure~\ref{fig:densfig}b.

From the data presented in Figure~\ref{fig:densfig}, we can conclude that 
the unbalanced forces arising from LJ attractions are the driving force for drying at extended hydrophobic interfaces. 
Indeed, we have previously shown that the net force on a water molecule at an extended hydrophobic interface
from long ranged electrostatics is much smaller than that from LJ attractions~\cite{JStatPhys}.
Nevertheless, long ranged electrostatics play a subtle but important role
in determining the orientational preferences of water
and properties dependent upon this orientational structure.
One such quantity is the electrostatic or polarization potential $\Phi(r)$ felt
by a test charge
\begin{eqnarray}
\Phi(r)&=&-\int_0^r dr' \E(r') \nonumber \\
&=&-\int_0^r\frac{dr'}{r'^2}\int_0^{r'}dr''r''^2\rhoq(r''),
\label{eq:potprof}
\end{eqnarray}
where $\rhoq(\rb)\equiv \avg{\sum_i q_i \delta(\rb-\rb_i)}$ is the ensemble averaged
charge density of the system and $\E(r)$ is the electric field due
to the polarization of water molecules induced by the presence of the solute.

\begin{figure}[tb]
\centering
\includegraphics{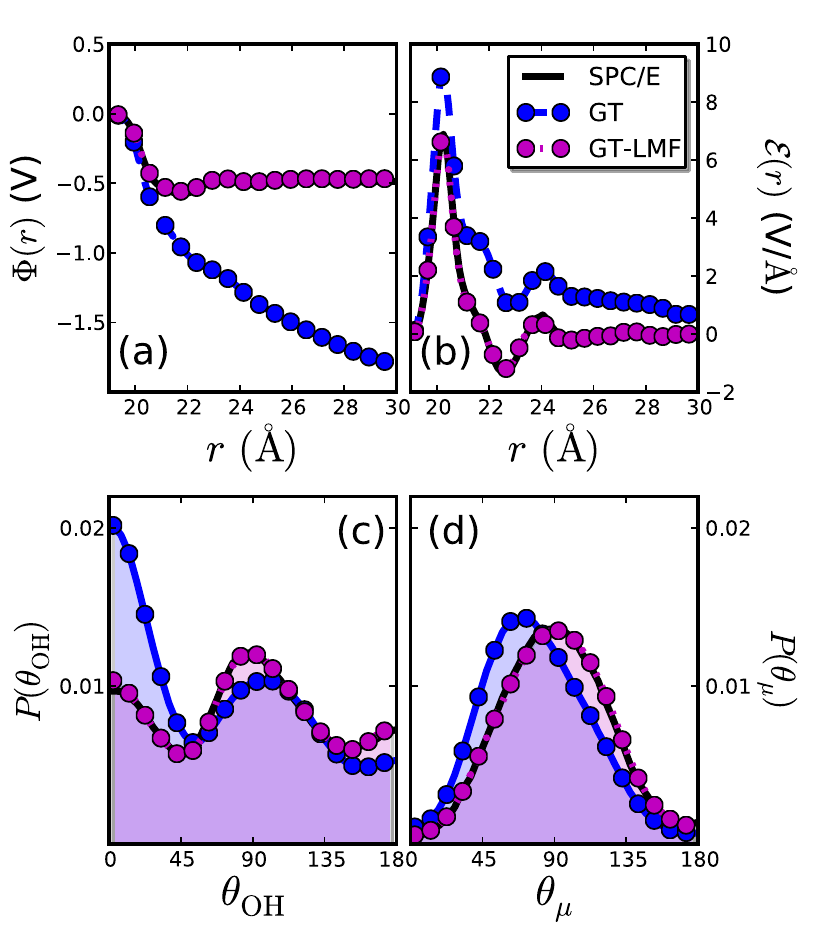} 
\caption{(a) Polarization potential $\Phi(r)$ and (b) the corresponding electric fields $\E(r)$
obtained for a solute of $\RHS\approx20$~\AA \ in SPC/E and
GT water, as well as for GT water in the presence of the electrostatic LMF (GT-LMF).
(c) Probability distributions of the angle formed by the OH bond vector and the vector connecting the oxygen
site with the center of the solute ($\toh$), for molecules within 1~\AA \ of the solute surface for the three systems
shown in (a). The analogous distributions for the dipolar angle $\tmu$ are shown in (d).}
\label{fig:elec}
\end{figure}

The polarization potential of SPC/E water, shown in
Figure~\ref{fig:elec}a, reaches a constant value of
approximately 500~mV in the bulk region,
consistent with previous determinations of interface potentials at extended hydrophobic interfaces
for this water model~\cite{LMFWater}.
Removal of the long ranged electrostatic interactions in GT water leads to an approximate charge density that
does not predict this plateau in the bulk region, Figure~\ref{fig:elec}a. Thus there is
a net electric field $\E(r)$ in this system, even far from the solute surface as shown in Figure~\ref{fig:elec}b.
The appearance of a non-vanishing electric field in the bulk of GT water
is associated with an over-orientation of interfacial OH bonds toward
the solute surface.
This is evidenced by a larger peak at $\toh\approx0\degree$
in the probability distribution $P\para{\toh}$ for interfacial GT water molecules
in comparison to that observed for SPC/E water,
shown in Figure~\ref{fig:elec}c,
where $\toh$ is the angle formed
by the OH bond vector and the oxygen-solute vector

The increase in the number of OH groups pointing toward the
interface in GT water is driven by the tendency to maintain the hydrogen bond network alone. 
This results in the formation
of an overly ordered dipole layer at the interface, demonstrated by the peak at 
$\tmu\approx60\degree$ in $P\para{\tmu}$, shown in Figure~\ref{fig:elec}d,
where $\tmu$ is the angle formed
by the dipole vector of water and the oxygen-solute vector.
Without long ranged dipole-dipole interactions, water far from the surface
does not respond to the presence of this dipole layer, and
$\E(r)$ remains non-zero well into the bulk region. However, we can compensate for the
averaged effects of the long ranged electrostatics through the introduction of the electrostatic LMF
for an uncharged solute~\cite{LMFDeriv}
\begin{equation}
\Vr(\rb)=\int d\rb' \rhoqr(\rb')v_1\para{\len{\rb-\rb'}},
\label{eq:clmf}
\end{equation}
where $v_1(r)=\erf(r/\sigma)/r$ is the long ranged, slowly varying component of $1/r$, separated
with a smoothing length $\sigma=4.5$~\AA~\cite{JStatPhys} herein, and
in general $\sigma$ should be chosen to be greater than the nearest-neighbor distance in a fluid~\cite{LMFDeriv}.
Inclusion of this renormalized solute potential in the GT water system
leads to quantitative accuracy of both the electrostatic and orientational structure of interfacial water,
evidenced by the curves labeled GT-LMF in Figure~\ref{fig:elec}.

%@@@@@@@@@@@@@@@@@@@@@@@@@@@@@@@@@@@@@@@@@@@@@@
%@@@@@@@@@@@@@@@@@@@@@@@@@@@@@@@@@@@@@@@@@@@@@@

In his seminal work on nonpolar solutes in
aqueous solutions, Stillinger
deduced that orienting an OH bond toward the interface
provides the least energetic detriment to the hydrogen bond network of water \cite{StillingerLV}.
In GT water there are no opposing long ranged electrostatic interactions and the energetics of the
hydrogen bond network alone determines the orientational preferences of water at the interface. 
However, this results in too high a probability
of pointing an OH bond toward the interface, illustrating that while hydrogen bonding is
a major driving force in determining the structure of water around large apolar solutes,
it is not the sole determinant of the observed orientational preferences of interfacial water.

In an earlier contribution, Stillinger and Ben-Naim initially postulated
that the dipole and quadrupole moments of water lead
to a mean torque on a molecule at the interface with its vapor
that orients the dipole moment of an interfacial water molecule
toward the bulk liquid~\cite{SBNPotentialDrop}. 
This behavior is reflected in the change of $P\para{\tmu}$ upon the inclusion of long ranged
interactions through $\Vr$,
which provides the slowly-varying torque necessary to slightly turn the molecular dipoles of interfacial water 
in the direction of the bulk and obtain the desired orientational structure, evidenced by the distributions
$P\para{\tmu}$ shown in Figure~\ref{fig:elec}d. 
Therefore, the orientational structure
of water at extended hydrophobic surfaces is a result of a
delicate balance of the energetics of the hydrogen
bond network \textit{and} the mutipolar interactions arising from the electrical asymmetry
of a water molecule, with the former dominating.

\section{The Response of Interfacial Water to Unbalanced Forces}

In this section, we examine the response of short ranged reference systems around solutes of 
varying sizes to the presence of very strong unbalanced forces
like those seen in reality only for very large solutes. 
This provides a stringent test of the stability of the hydrogen bond network 
around small solutes even when subjected to strong perturbations.
In order to accomplish this task, we scale the long ranged LJ LMF determined 
for a large solute of radius $\RHS\approx20$~\AA \ by its radius, and then rescale the field to the 
desired solute size, $\RHStilde$,
\begin{equation}
\phiRltilde\para{r;\lambda,\RHStilde}=\lambda\phiRlLJ\para{\frac{\RHStilde}{\RHS}r;\RHS}.
\end{equation}
where $\phiRlLJ(\rb)$ is the slowly-varying renormalized portion of the LMF
shown in the inset of Figure~\ref{fig:densfig}b.
Here the notation $\phiRlLJ\para{r;\RHS}$ indicates that the field
$\phiRlLJ$ is a function of $r$ and that it was
determined when a solute of radius $\RHS$ is fixed at the origin.
The fictitious, rescaled LMF is indicated by $\phiRltilde$, and the coupling parameter $\lambda$ is used to 
further adjust the magnitude of this field.
In effect we have taken the large unbalanced LJ force around a large solute, 
which Figure~\ref{fig:densfig}b shows is strong enough to 
significantly perturb the large scale density profile of GTRC water when corrected with LMF theory, 
and artificially applied it to a small scale system like that in Figure~\ref{fig:densfig}a
with an intact local hydrogen bond network.
This provides insight into the very different response interfaces
around small and large hydrophobic solutes have to repulsive forces over a wide range of magnitude as
$\lambda$ is varied, including exceptionally large unbalanced forces seen in reality only near large
hydrophobic solutes.

In order to quantify the response of water to strong unbalanced forces, 
we focus on the $\lambda$-dependence of the average number of water molecules in the
solute solvation shell, 
$\avg{N(\lambda)}_{\phiRltilde}$,
as well as the corresponding response function
\begin{equation}
\chi(\lambda)=-\frac{1}{\avg{N(0)}_{\phiRltilde}}\para{\frac{\partial \avg{N(\lambda)}_{\phiRltilde}}{\partial \lambda}},
\end{equation}
where $\avg{\cdots}_{\phiRltilde}$ indicates that the ensemble average
is performed in the presence of the field $\phiRltilde\para{r;\lambda,\RHStilde}$. 
The function $\avg{N(\lambda)}_{\phiRltilde}$ is calculated for distances
$r<r_{\rm min}$, where $r_{\rm min}$ is defined as the distance at which
the density distribution in the absence
of the field reaches its first minimum.

%@@@@@@@@@@@@@@@@@@@@@@@@@@@@@@@@@@@@@@@@@@@@@@
\begin{figure}[tb]
\centering
\includegraphics{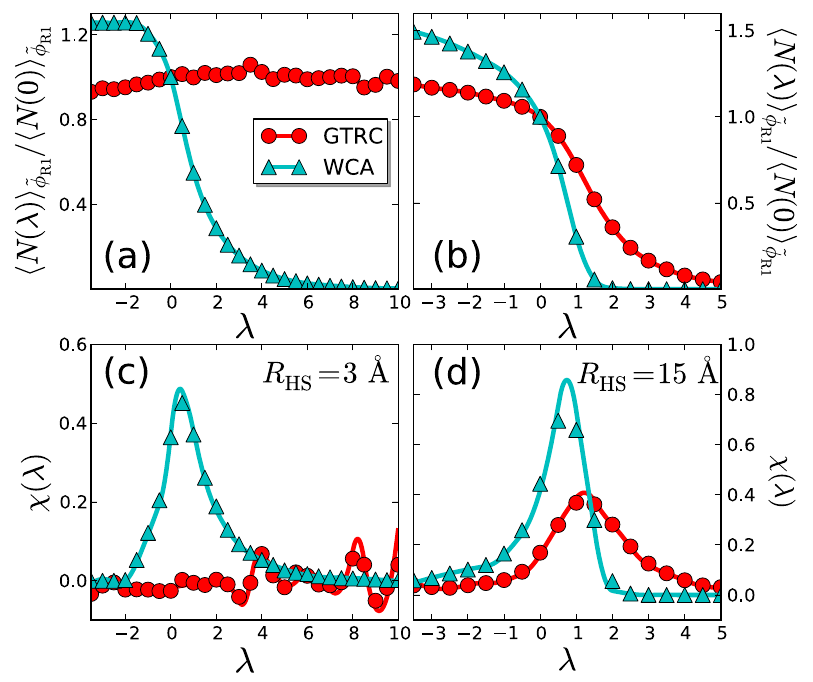} 
\caption{Average number of truncated water and LJ molecules in the first solvation shell as a function of the coupling parameter
$\lambda$ for solutes of radii $\RHS\approx3$~\AA \ (a) and $\RHS\approx15$~\AA \ (b).
Results are shown for both GTRC water and 
the WCA fluid (with the LJ parameters of SPC/E water), 
and $\avg{N(\lambda)}_{\phiRltilde}$ has been normalized by its value in the case of zero
field in order to make comparisons between the two fluids.  The corresponding
response functions are shown in (c) and (d), respectively.  Solid lines in (a) and (b) are spline fits
to $\avg{N(\lambda)}_{\phiRltilde}$ and those in (c) and (d) are the negative derivatives of the corresponding fits.}
\label{fig:response}
\end{figure}
%@@@@@@@@@@@@@@@@@@@@@@@@@@@@@@@@@@@@@@@@@@@@@@

In the large scale hydration regime the broken hydrogen bonds in the
interfacial region effectively permit the interface to detatch from the solute and the interface
is ``soft'' and fluctuating. We expect water to have a response qualitatively similar to that
of simple liquids where drying occurs with increasing strength of $\phiRltilde$. 
However, in the small length scale limit, while network fluctuations certainly occur, the hydrogen bond network is
basically maintained around the solute.
We thus expect that the small scale solute-water interface is ``stiff''
and highly resistant to perturbations unless they are strong enough to break hydrogen bonds.
This should lead to behavior that is fundamentally different from that of a simple LJ fluid,
which lacks such strong, local interactions. 

As postulated above, in the large solute regime, 
the behavior of $\avg{N(\lambda)}_{\phiRltilde}$ and $\chi(\lambda)$ are qualitatively similar for both
GTRC water and the WCA fluid (Figure~\ref{fig:response}b and Figure~\ref{fig:response}d). 
Gradual dewetting is observed with increasing field strength, until no molecules are present
in the solvation shell region at high values of the coupling parameter.
In fact, as $\lambda$ is increased, a peak in the response function $\chi$ is observed, 
indicative of a drying transition in the hydration shell of the solute; the details of the transition
differ between GTRC water and the WCA fluid due to differences in state points and interaction potentials.

In the small solute regime, the WCA fluid displays 
signatures of a drying transition completely analogous to those
seen in the large solute case with a simple shift in $\lambda$.
GTRC water, on the other hand, does not display characteristics of such
nanoscale dewetting (Figure~\ref{fig:response}a and Figure~\ref{fig:response}c);
$\avg{N(\lambda)}_{\phiRltilde}$ stays roughly constant and
the response function fluctuates about zero.
Using a typical geometric definition of a hydrogen bond~\cite{LuzarChandlerPRLHbondDef,JStatPhys},
we find that the average number of hydrogen bonds per molecule,
for waters located between the solute and
the position of the first maximum in the corresponding $\rho(r)$,
fluctuates around 3.5 for all $\lambda\ge0$, very close to the bulk
value of 3.6 hydrogen bonds per water molecule.
Therefore, the hydrogen bond network is maintained around
the small solute for all studied values of $\lambda$, and 
the strong local interactions of the hydrogen bond network
prohibit drying at the solute surface, even
in the presence of the extremely large external fields considered herein.

The above-described results indicate that the underlying physics behind the solvation behavior in a LJ fluid
is qualitatively similar in the small and large length scale regimes, 
dependent only on the magnitude of the unbalancing
potential arising from the bulk, while that of water qualitatively
differs in the two regimes. In the large length scale regime,
water behaves in a manner similar to a LJ fluid,
with the unbalanced LJ attractions having a substantial impact on the
solvation structure. For solutes smaller than the crossover radius,
however, water wets the surface of the solute even
in the presence of extremely large (though fictitious) unbalancing potentials;
the hydration shell remains intact due
to the great strength of the local hydrogen bond network. Therefore,
interfacial fluctuations and the physics dictating where the length scale transition occurs
is different for water than for simple, non-hydrogen bonding fluids.

\section{Hydrogen bonding sets the scale for the crossover in hydration thermodynamics}
\begin{figure}[tb]
\centering
\includegraphics{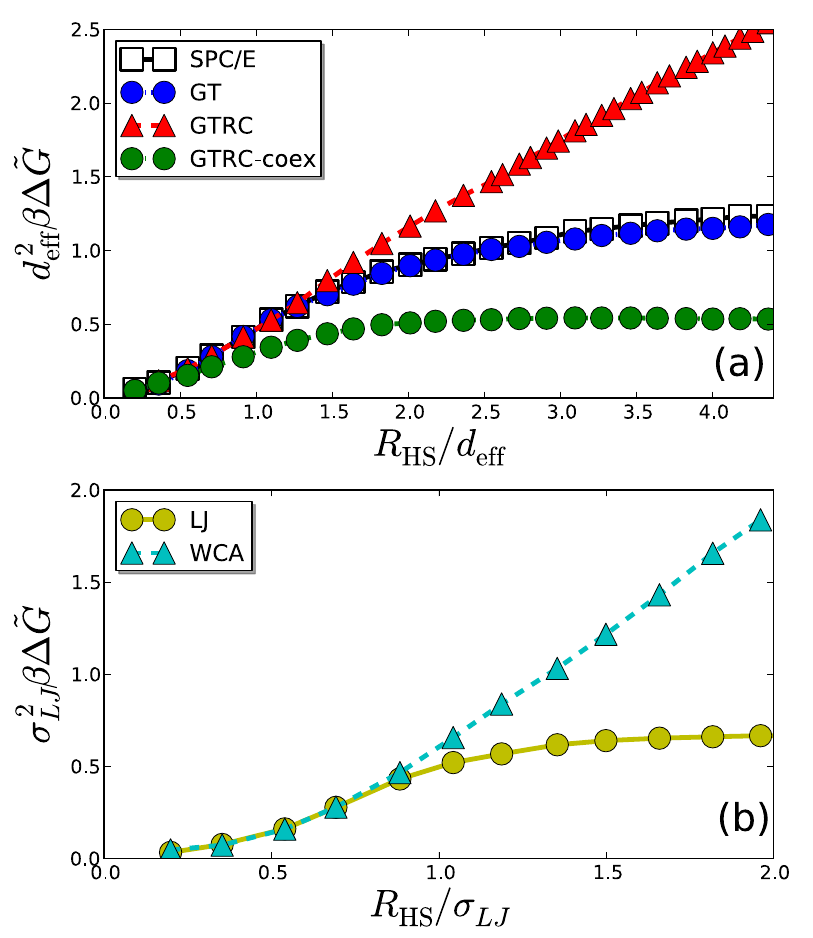} 
\caption{Solvation free energies of apolar spheres per unit solute area as a function of solute radius,
scaled by the effective diameter of a corresponding solvent molecule
($d_{\rm eff} = 2.75$~\AA \ for water),
for (a) SPC/E, GT, and GTRC
water models, as well as (b) a LJ fluid and its corresponding WCA reference system. 
Error bars are smaller than the symbols shown.}
\label{fig:sfe}
\end{figure}

The above-described physical balance between hydrogen bonding and interfacial unbalancing potentials
also plays a key role in the solvation thermodynamics of apolar solutes.
Gibbs free energies of solvation, $\Delta G$,
were calculated by performing equilibrium simulations of solutes with 
effective hard sphere radii $\RHS\le13$~\AA \ in increments of $\Delta\RHS\approx0.5$~\AA.
Due to poor phase space overlap between neighboring windows, $\Delta\RHS$ was
decreased to $0.25$~\AA \ to determine $\Delta G$ for solutes with $\RHS>7$~\AA \
solvated by GTRC water.
The solvation free energies presented herein were calculated using
the Bennett acceptance ratio or BAR~\cite{BAR,GoodPractices} method.
To emphasize the crossover in the scaling behavior of the solvation free energies,
we normalize $\Delta G$ by the surface area of the apolar solute (Figure~\ref{fig:sfe}),
$\Delta\tilde{G}=\Delta G/4\pi\RHS^2$.

In the small solute regime, $\RHS\le R_C\approx5.0$~\AA, the hydration free energies
are in agreement for all three models. This illustrates that the hydration
thermodynamics of small, nonpolar solutes are dictated by the local structure
of water alone, as would be expected from the conclusions drawn above regarding
solvation in the SPC/E, GT, and GTRC models.
Indeed, the dominant role of local structure in the small solute regime is not restricted to water, as indicated by the agreement of the
solvation free energies for LJ and WCA fluids for small solute sizes
shown in Figure~\ref{fig:sfe}b. 

The free energy for large solutes scales with surface area in both SPC/E water and the LJ fluid,
and here long ranged interactions become increasingly important. Only small differences
in $\Delta G$ are observed between SPC/E and GT water,
reflecting the relatively small role of long ranged electrostatics in hydrophobic hydration~\cite{JStatPhys}.
LJ attractions, on the other hand, make a substantial contribution to the hydration free energy.
Indeed because of the absence of these attractions, GTRC water
completely lacks the plateau in $\Delta\tilde{G}$ for large solute sizes.

The behavior of the GTRC water model can be explained by noting that in the large solute regime,
$\Delta G\sim PV_S + \gamma A_S$, where $V_S$ and $A_S$
are the volume and surface area of the solute, respectively, $P$ is the pressure of the system, and
$\gamma$ is the solute-water surface tension. In order to obtain the same bulk density as SPC/E water
at a pressure of 1~atm, the GTRC model must be maintained at a pressure of roughly 3~katm. At this state
the GTRC water model is far from liquid-vapor coexistence, and the pressure is large enough to make the $PV_S$ term dominate the 
behavior of $\Delta G$ for large solutes.  

%@@@@@@@@@@@@@@@@@@@@@@@@@@@@@@@@@@@@@@@@@@@@@@
%@@@@@@@@@@@@@@@@@@@@@@@@@@@@@@@@@@@@@@@@@@@@@@

However we have previously
shown that GTRC water can indeed have a self-maintained liquid-vapor interface,
but at a lower bulk density close to that of ice. The interface is maintained by the
strong short ranged Coulomb attractive forces between donor and acceptor sites and the need to preserve as many hydrogen bonds as possible~\cite{JStatPhys}. However, because there are no unbalanced forces
from LJ attractions, the surface tension is much smaller than that of the full SPC/E model.

As shown in the curve labeled `GTRC-coex' in Figure~\ref{fig:sfe}a,
the solvation free energies in GTRC water near coexistence in both the small and large solute regimes are smaller in magnitude
than those in SPC/E water. However it exhibits essentially the same crossover radius as the full SPC/E model and
scales with solute surface area for large solutes.
The behavior of $\Delta G$ below the crossover radius can be understood from our previous results for the
bulk structure of the GTRC model near coexistence~\cite{JStatPhys}. The bulk coexistence density is close to that of ice and the hydrogen bond network has a more ordered tetrahedral structure that can more readily accommodate the formation of a cavity than is the case for SPC/E water.

Although the solvation free energies of apolar solutes in water and in the LJ fluid
exhibit qualitatively similar crossover behavior, they differ in one important respect:
the length scale at which the crossover in solvation behavior occurs.
For the LJ fluid, the crossover radius is approximately equal to the diameter of a solvent particle. 
At this solute size, the unbalanced forces from the LJ attractions of the bulk region
become large enough to ``pull'' particles away from the solute surface,
leading to drying.

Although unbalanced LJ forces also exist when apolar particles
of similar size are solvated by water, the possible disruption of strong local hydrogen bonds between interfacial water
molecules dominates the energetics,  and the crossover occurs only when water is not able to maintain this network. This leads to an estimate for the crossover radius, $R_C\approx5$~\AA,
almost twice the diameter of a water molecule (2.75~\AA)
and significantly larger than that found in a LJ fluid. As shown above, hydrophobic solvation in
GTRC water near coexistence also displays a crossover in its scaling behavior at
a value of $R_C$ essentially the same as that of the full SPC/E model.
Because GTRC water accounts only for the
hydrogen bond network, we can conclusively say that the 
crossover in solvation behavior is determined by the hydrogen bond network of water alone,
occurring when the solute size is increased to a point beyond which it is impossible for this
network to remain intact, consistent with the original arguments of Stillinger \cite{StillingerLV}.

Given the importance of the hydrogen bond network for small scale solvation in water, how can we
rationalize the success of the LCW theory~\cite{LCW}  and related lattice models incorporating similar physics~\cite{VarillyHphobLattice,ChandlerPolymerLattice}, 
which lack an explicit description of hydrogen bonds? These theories correctly describe the small scale physics
driven by Gaussian density fluctuations in the bulk solvent and the large scale physics dominated by the formation of
a vapor-like interface around a large repulsive solute. Effective parameters controlling the transition between
the two regimes are fit to experimental data for each particular solvent.

The key experimental parameters 
determining the transition length scale in the LCW theory are the liquid-vapor surface tension,
and the bulk density and compressibility. The small compressibility and large surface tension of water compared to a LJ fluid implicitly 
accounts for the strength of the hydrogen bond network in bulk water
and the difficulty of disrupting it by interface 
formation for large solutes.
This allows the LCW theory to qualitatively describe the different transition
length scales in both water and a
LJ fluid~\cite{HuangChandlerPRE} using the same basic framework.
But LCW theory uses mean field ideas and square gradient and
other approximations, and errors are seen in its detailed predictions for certain
other properties like the interface
width~\cite{HuangChandlerPRE}. 
More detailed approaches describing structure and
fluctuations in both small and large length scale regimes are needed for quantitative calculations.

More recent work by Rajamani, Truskett, and Garde~\cite{GardeCrossover} has clarified the relation between bulk thermodynamics and the crossover radius. They suggested
that the crossover radius is proportional to the Egelstaff-Widom
length scale $l_{\rm EW}=\gamma\kappa_T$,
the product of the liquid-vapor surface tension $\gamma$ and the isothermal
compressibility $\kappa_T$~\cite{EgelstaffWidom}. Quantitative agreement can be achieved 
by using a microscopic compressibility that depends on the solute volume rather than the long wavelength bulk compressibility in conjunction with the solute-water interfacial tension
to estimate the crossover radius $R_C$.

A simple but stringent test of this idea is to compare the Egelstaff-Widom length scale of GTRC water
near liquid-vapor coexistence to that of SPC/E water. As discussed above, the crossover radius in GTRC water is essentially the same as in SPC/E water.  This is easily rationalized from our microscopic understanding of the very similar behavior of the hydrogen bond network around the solute in GTRC and SPC/E water. If this simple physics is reflected in the Egelstaff-Widom length scale, this too should be nearly the same although both the surface tension and bulk compressibility differ considerably in the two models.

Indeed, the compressibility $\kappa_T^{\rm GTRC}$ of GTRC water at $T=300$~K and a pressure of 1~atm is 0.087~katm$^{-1}$, roughly a factor of two larger than that of SPC/E at the same state point, 0.045~katm$^{-1}$, while the surface tension of the GTRC model $\gamma^{\rm GTRC} \approx27$~mN/m, is about half of that of the SPC/E model $\gamma^{\rm SPC/E}\approx54.7$~mN/m.
Here the value for SPC/E water was taken from the work of Sedlmeier and Netz~\cite{NetzTolmanLength} and
the surface tension of GTRC water was estimated by extrapolating the solvation free energies $\Delta \tilde{G}(R_{\rm HS})$
presented in Section~V to the limit $R_{\rm HS}\rightarrow\infty$. Thus, the Egelstaff-Widom length scales of SPC/E and GTRC
water are nearly equal, $l_{\rm EW}^{\rm SPC/E}=0.24$~\AA \
and $l_{\rm EW}^{\rm GTRC}=0.23$~\AA, respectively, as expected.

\section{Entropy convergence is a consequence of the hydrogen bond network}

%@@@@@@@@@@@@@@@@@@@@@@@@@@@@@@@@@@@@@@@@@@@@@@
\begin{figure}[tb]
\centering
\includegraphics{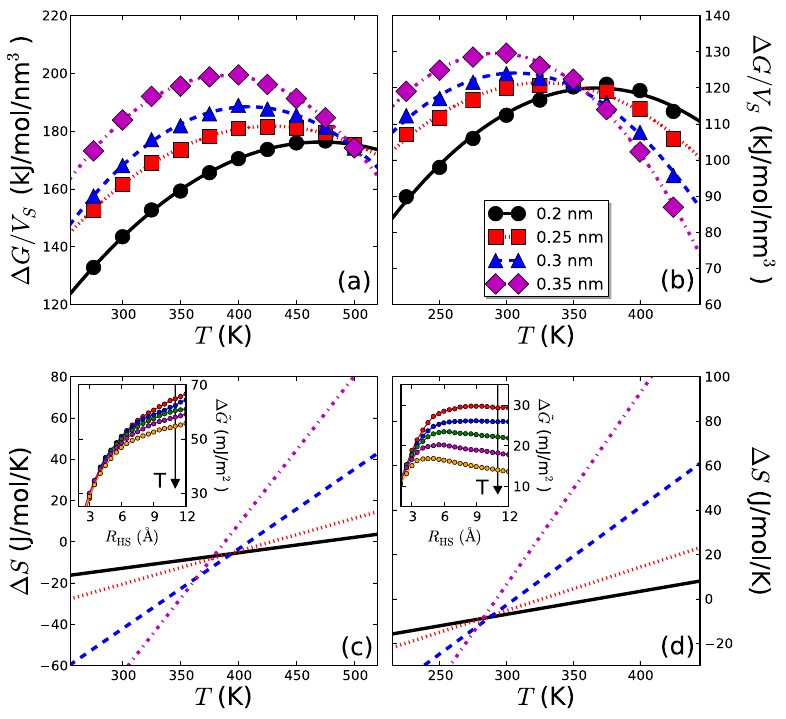} 
\caption{Hard sphere solvation free energy $\Delta G$ per unit solute volume $V_S$
as a function of temperature in (a) SPC/E and (b) GTRC water. The corresponding
entropies of solvation $\Delta S$ as a function of $T$ are shown in (c) and (d), respectively.
Hard sphere radii are indicated in the legend. Solvation free energies
as a function of solute size for $T=300$~K, 325~K, 350~K, 375~K, and 400~K are
shown in the insets of (c) and (d) for the SPC/E and GTRC models, respectively.
The arrows point in the direction of increasing temperature.}
\label{fig:entropy}
\end{figure}
%@@@@@@@@@@@@@@@@@@@@@@@@@@@@@@@@@@@@@@@@@@@@@@

The temperature dependence of hydrophobic hydration also displays features distinct from
solvation in typical van der Waals liquids. Specifically, hydration free energies $\Delta G$ of small
apolar particles increase with increasing temperature along a significant portion of the coexistence curve
until a maximum is reached. Above this temperature, free energies of solvation decrease with increasing temperature,
a behavior typical of most fluids. Associated with this region of anomalous solvation is the phenomenon of \emph{entropy convergence},
in which the hydration entropies,
$\Delta S = -(\partial \Delta G/\partial T)_P$, intersect near a temperature of $400$~K for a large range of solute sizes,
although the location of the hydration free energy maximum varies somewhat with solute size.
Analogous to the discussion of the crossover length scale, the explanation of entropy convergence typically
uses thermodynamic arguments, citing the small and nearly constant compressibility
of water along the liquid-vapor coexistence line, relative to organic solvents~\cite{GardeEntropy,HummerReview,NetzEntropy},
although explanations exist that do not hinge on the relative incompressibility of bulk water~\cite{GrazianoVDW}. 

In this section, we show that entropy convergence in water arises from the
hydrogen bond network through its impact on bulk thermodynamics
by studying the temperature dependence of hard sphere solvation in the SPC/E and GTRC water
models near liquid-vapor coexistence. Simulations of bulk SPC/E and GTRC water were carried
out at a pressure of 1~atm and
temperatures ranging from 275K--500K and 225K--425K, respectively. Hard sphere solvation
free energies in the small solute regime were determined by assuming Gaussian bulk density fluctuations~\cite{HummerInfoTheory,HummerReview,GardeEntropy}, 
\begin{equation}
\Delta G\approx\frac{\kT\rhob^2(T) V_S^2}{2\sigma_{V_S}(T)}+\frac{\kT}{2}\ln\brac{2\pi\sigma_{V_S}(T)},
\label{eq:Gaussian}
\end{equation}
where $\sigma_{V_S}=\avg{(\delta N)^2}_{V_S}$
is the mean squared fluctuation in the number of molecules $N$ in a solute-sized probe volume $V_S$, with
$\delta N=N-\avg{N}_{V_S}$, and we consider the volume $V_S=4\pi R_{\rm HS}^3/3$ of a spherical solute of radius
$R_{\rm HS}$ herein.
These solvation free energies were then fit to 
$\Delta G(T)=a+bT-cT^2$, and are plotted as lines in Figures~\ref{fig:entropy}a and~\ref{fig:entropy}b.
Solvation entropies were determined from the negative derivative of these
fits, and are shown in Figures~\ref{fig:entropy}c and~\ref{fig:entropy}d.

The temperature dependence of hard sphere solvation is qualitatively similar in both SPC/E
and GTRC water. In fact, entropy convergence is observed in the GTRC model, albeit 
at a convergence temperature $\tilde{T}$ approximately 100~K less than the convergence temperature in SPC/E water;
$\tilde{T}_{\rm SPC/E}=387\pm8$~K and $\tilde{T}_{\rm GTRC}=291\pm7$~K, obtained from linear fitting
of $\Delta S$ as a function of the heat capacity of solvation, $\Delta C_P(T)=T(\partial \Delta S/\partial T)_P$,
for several temperatures~\cite{NetzEntropy}.
Despite this quantitative distinction,
the fact that the minimal reference network of GTRC water captures the phenomena of
entropy convergence explicitly demonstrates that this signature of hydrophobic hydration
is directly related to the energetics of the hydrogen bond network over a wide range of temperatures.

Previous work has shown that the logarithmic term in Equation~\ref{eq:Gaussian} has merely a secondary effect on entropy
convergence, shifting $\tilde{T}$ to somewhat lower values and $\Delta S(\tilde{T})$ from zero to negative values~\cite{GardeEntropy}.
Therefore, in order
to obtain a qualitative, microscopic explanation for entropy convergence, we can 
neglect this term in the Gaussian approximation for the free energy, and write the solvation entropy as
\begin{equation}
\Delta S \approx -\para{\frac{k_B V_S^2}{2\sigma_{V_S}}}\rhob^2(T)\brac{1-2T\alpha_P(T)},
\label{eq:ds}
\end{equation}
where $\alpha_P=-(\partial \ln\rhob/\partial T)_P$ is the thermal expansion coefficient at constant pressure, which was
determined by fitting the bulk densities to Laurent polynomials~\cite{AshbaughTMD}.
Here we have also assumed that the temperature dependence of the variance $\sigma_{V_S}$ can be neglected, as has been previously established~\cite{GardeEntropy}.
Thus within the accuracy of Equation~\ref{eq:ds}, entropy convergence is seen for
$\Delta S(\tilde{T})=0$, and an estimate of the convergence temperature
can be obtained from the intersection of $\alpha_P(T)$ and $(2T)^{-1}$. 
The convergence temperatures obtained for the SPC/E and GTRC
models from Equation~\ref{eq:ds} are roughly 420 K and 330 K, respectively,
in reasonably good agreement with the results presented above, although $\tilde{T}$ will always be overestimated in this approximation.
Nonetheless, the difference between the convergence temperatures
of the two models is quantitatively captured by this estimation,
indicating that additional $T$-dependences arising in $\Delta G$ are
similar in the two models, and these have been discussed in detail elsewhere~\cite{GardeEntropy,NetzEntropy}.

%@@@@@@@@@@@@@@@@@@@@@@@@@@@@@@@@@@@@@@@@@@@@@@
\begin{figure}[tb]
\centering
\includegraphics{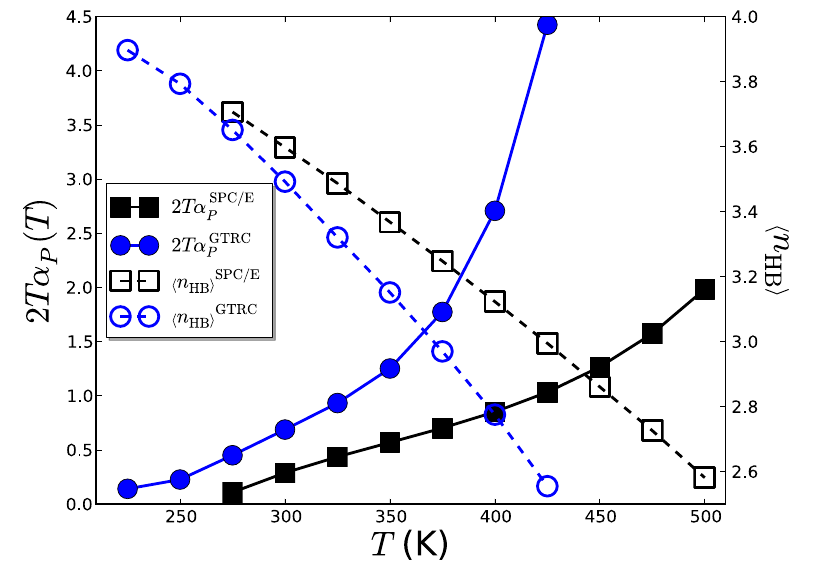} 
\caption{Thermal expansion coefficient multiplied by twice the temperature
(left axis, closed symbols) and average number of hydrogen bonds
per water molecule (right axis, open symbols) for SPC/E and GTRC water.}
\label{fig:alpha}
\end{figure}
%@@@@@@@@@@@@@@@@@@@@@@@@@@@@@@@@@@@@@@@@@@@@@@

In this simplified Gaussian framework the behavior of $\alpha_P(T)$ plays a key role in entropy convergence.
In the case of SPC/E water, the thermal expansion coefficient vanishes at 
the temperature of maximum density near 248~K~\cite{AshbaughTMD,JStatPhys}. As shown in Figure~\ref{fig:alpha},
$2T\alpha_P(T)$ then increases with increasing temperature but remains less than one until about 420 K,
where entropy convergence is predicted to occur.
The thermal expansion coefficient of GTRC water behaves in a qualitatively similar manner with $2T\alpha_P(T)$ remaining less than one until about 330K, although $\alpha_P(T)$ is never negative, because
this model lacks a density maximum near liquid-vapor coexistence~\cite{JStatPhys}.  The behavior of the thermal expansion coefficient
is a direct consequence of the energetics of the H-bond network in both models. At ambient temperatures, the average number of hydrogen bonds
per molecule approaches four in both SPC/E and GTRC water~\cite{JStatPhys}. With increasing temperature, thermal fluctuations
increasingly disrupt the entropically unfavorable hydrogen bond network in both models (Figure~\ref{fig:alpha}),
which leads to an increase in the thermal expansion coefficient.
However, the lower density of GTRC water
permits more fluctuations as the temperature is increased, consistent with
its larger compressibility and a more rapid increase in $\alpha_P(T)$, leading to a lower convergence temperature.

We also determined the temperature dependence of large solute solvation free energies
following the description in the previous section. After the length scale transition, solvation is dominated by
interfacial physics. As evidenced by the insets in Figure~\ref{fig:entropy},
hard sphere solvation free energies in this regime
decrease with increasing temperature for both models, following the $T$-dependence
of the surface tension, just as is the case for LJ solvation.

\section{Long ranged interactions and the size dependence of hydrophobic association}

In this section, we examine the role of the various short and long ranged forces
in the thermodynamics of hydrophobic association. In order to accomplish this task,
we consider the association of pairs of spherical solutes, one pair in which both solutes
are in the small-scale regime, while the other pair consists of two large solutes.
We first examine the free energy as a function of solute-solute distance, $R$,
\begin{equation}
\beta W(R)=-\ln P(R),
\end{equation}
where $P(R)$ was obtained by umbrella sampling with the harmonic biasing potential 
\begin{equation}
U_{\rm bias}(R)=\frac{\kappa}{2}\para{R-R^*}^2,
\label{eq:rumb}
\end{equation}
$R^*$ is the desired value of $R$, and $\kappa$ is a force constant
tuned to achieve adequate overlap between neighboring windows.
The probability distribution $P(R)$ was then
constructed from the set of biased simulations using the
multistate Bennet acceptance ratio method (MBAR)~\cite{MBAR}.

We first focus on hydrophobic association in the small scale regime, and consider
the association of two united atom (UA) methane models, which are simply
LJ particles with length and energy parameters of $\sigma_{\rm Me-Me}=3.73$~\AA \
and $\epsilon_{\rm Me-Me}=1.234$~kJ/mol, respectively~\cite{MethaneParms}. Methane-water interactions
were obtained from Lorentz-Berthelot mixing rules.

%@@@@@@@@@@@@@@@@@@@@@@@@@@@@@@@@@@@@@@@@@@@@@@
\begin{figure}
\centering
\includegraphics{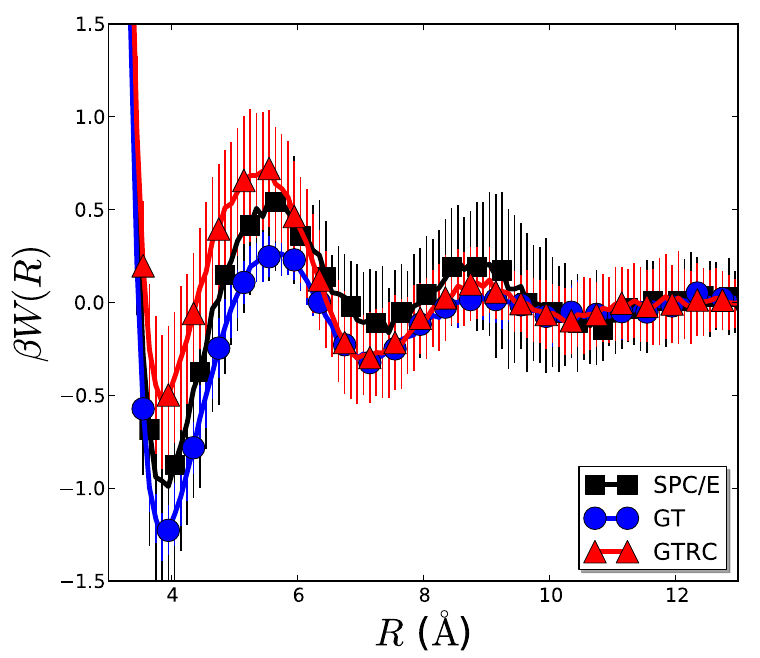} 
\caption{Potential of mean force, $W(r)$, between two UA methane particles in
SPC/E, GT, and GTRC water.}
\label{fig:mepmf}
\end{figure}
%@@@@@@@@@@@@@@@@@@@@@@@@@@@@@@@@@@@@@@@@@@@@@@

The potentials of mean force, $W(R)$, shown in Figure~\ref{fig:mepmf}
for the association of two UA methanes are nearly identical for all water models under
consideration. Therefore, not only does the hydrogen bond network dictate the solvation
structure around individual small solutes, but also the association of solutes in this length scale regime,
as expected from the results presented in the previous sections.

We now consider the association of two large C60 fullerene molecules in the various models of water.
Each C60 is represented as a single site using the coarse-graining procedure
prescribed by Girifalco~\cite{Girifalco, AACGC60}, such that the fullerene-fullerene interaction is given by
\begin{eqnarray}
U_{\rm FF}(R)&=& -\alpha\brac{ \frac{1}{s(s-1)^3} + \frac{1}{s(s+1)^3} - \frac{2}{s^4} } \nonumber \\
&+&\zeta\brac{ \frac{1}{s(s-1)^9} + \frac{1}{s(s+1)^9} - \frac{2}{s^{10}} },
\end{eqnarray}
where $\alpha=4.4775$~kJ/mol, $\zeta=0.0081$~kJ/mol, $s=R/2\eta$, and $\eta=3.55$~\AA.
The C60-water interaction potential is
\begin{eqnarray}
U_{\rm wF}(r)&=& 4N\epsilon_{\rm wF}\frac{\sigma_{\rm wF}^2}{r\eta} \Bigg\{ \frac{1}{20}\brac{
\para{ \frac{\sigma_{\rm wF}}{\eta-r}}^{10} - \para{\frac{\sigma_{\rm wF}}{\eta+r}}^{10} } \nonumber \\
&-&\frac{1}{8}\brac{ \para{\frac{\sigma_{\rm wF}}{\eta-r}}^4 - \para{\frac{\sigma_{\rm wF}}{\eta+r}}^4 }\Bigg\},
\end{eqnarray}
where $N=60$, $\sigma_{\rm wF}=3.19$~\AA, and $\epsilon_{\rm wF}=0.392$~kJ/mol.
Previous work has shown that this coarse-grained water-C60 interaction provides a very
good representation of the solvation structure in the corresponding
atomically-detailed water-C60 system~\cite{AACGC60}.

The water-C60 interaction potential $U_{\rm wF}(r)$ leads to a hydrophilic particle due to the high
density of carbon sites on the surface of the C60 molecule. Therefore, we also consider a hydrophobic
particle obtained by using only the repulsive water-C60 and C60-C60 forces. This is obtained by performing
a WCA-like separation of the potentials $U_{\rm FF}$ and $U_{\rm wF}$ to obtain the corresponding
purely repulsive potentials $U_{0,\rm FF}$ and $U_{0,\rm wF}$, as detailed above for $U_{\rm sw}$.

%@@@@@@@@@@@@@@@@@@@@@@@@@@@@@@@@@@@@@@@@@@@@@@
\begin{figure}
\centering
\includegraphics{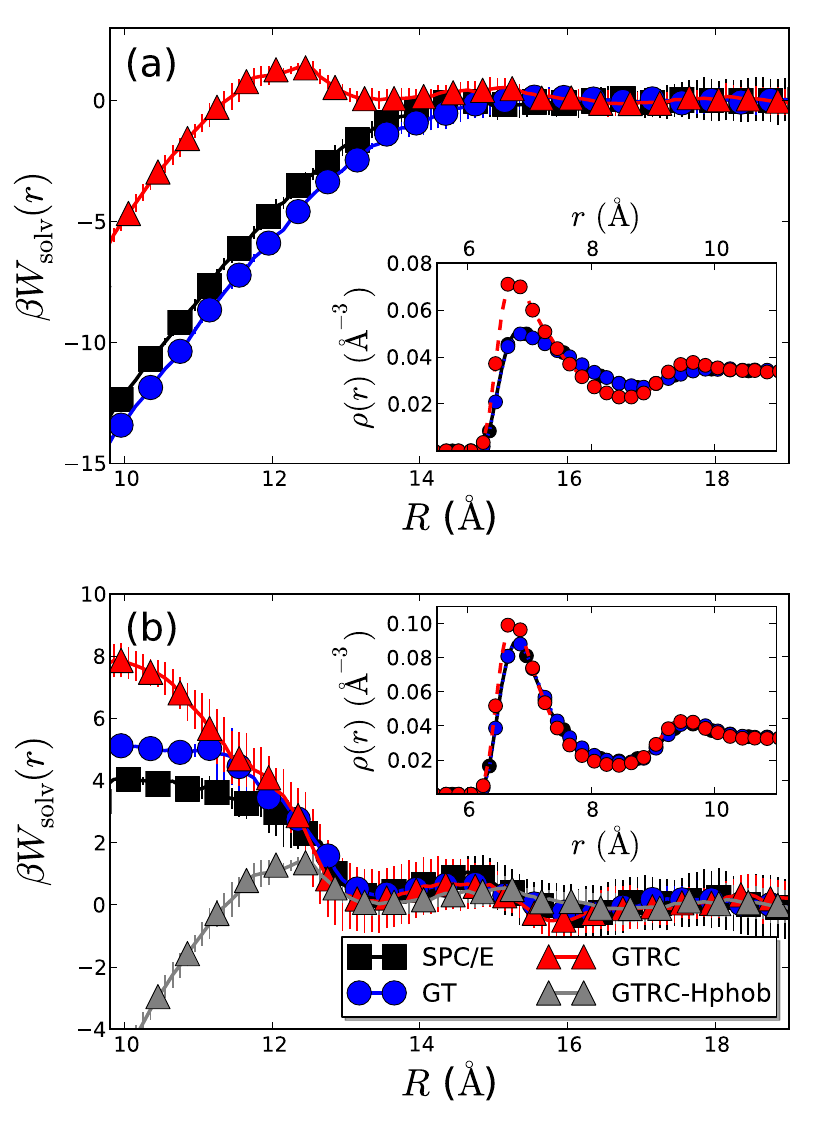} 
\caption{Solvent-induced potential of mean force, $W_{\rm solv}(R)$, between two (a) purely repulsive
and (b) attractive coarse-grained
C60 particles in SPC/E, GT, and GTRC water. Insets in (a) and (b) show the corresponding
nonuniform densities around a single coarse-grained C60 immersed in each water model.
The gray curve in (b) is the GTRC PMF from panel (a).}
\label{fig:c60pmf}
\end{figure}
%@@@@@@@@@@@@@@@@@@@@@@@@@@@@@@@@@@@@@@@@@@@@@@

We further separate the potential of mean force as $W(R)=W_{\rm vac}(R)+W_{\rm solv}(R)$, where
$W_{\rm vac}(R)$ and $W_{\rm solv}(R)$ are the vacuum and solvent-induced portions of the PMF,
respectively, focusing on the latter contribution herein.
The solvent-induced PMFs between purely repulsive
C60 particles in the SPC/E and GT water models, shown in Figure~\ref{fig:c60pmf}a,
are indicative of the hydrophobic
effect; the association of two large apolar particles in water is barrierless,
although the free energy of association is slightly lower in GT water due to its
lower surface tension.
Previous work has shown that the collapse of two extended hydrophobic surfaces proceeds
by the formation of a vapor tube~\cite{LuzarVaporTube, WillardChandlerCGSolutes},
in which solvent molecules are evacuated from
a cylindrical region between the two hydrophobes, and we will show below that the association
of two repulsive fullerenes also occurs by this mechanism.

In GTRC water, however, the PMF $W_{\rm solv}(R)$ displays a slight barrier
at $R\approx15$~\AA, and another significantly higher barrier at $R\approx12$~\AA, as shown in Figure~\ref{fig:c60pmf}a. 
Because the C60-C60 distance does not explicitly account for changes in the behavior of the aqueous
solvent, it is not a good reaction coordinate to study the association of two large hydrophobes on its
own~\cite{LuzarVaporTube, WillardChandlerCGSolutes}
and $W(R)$ cannot provide an explanation for the appearance of this barrier
in $W_{\rm solv}(R)$.

To understand hydrophobic association in GT and GTRC water, we calculate
the free energy as a function of the C60-C60 distance $R$ and the density $\rho_v$ of water
in a cylindrical volume of radius 3.75~\AA \ between the particles. 
This two-dimensional free energy landscape is given by
$\beta W(R,\rho_v)=-\ln P(R,\rho_v)$, where
$P(R,\rho_v)$ was calculated using the indirect umbrella sampling method~\cite{INDUS}
to bias the number of particles in
the volume $v$. The harmonic potential in Eq.~\ref{eq:rumb} was used to bias $R$.
Again MBAR was used to reconstruct
the probability distribution from these biased simulations~\cite{MBAR}.

\begin{figure}
\centering
\includegraphics{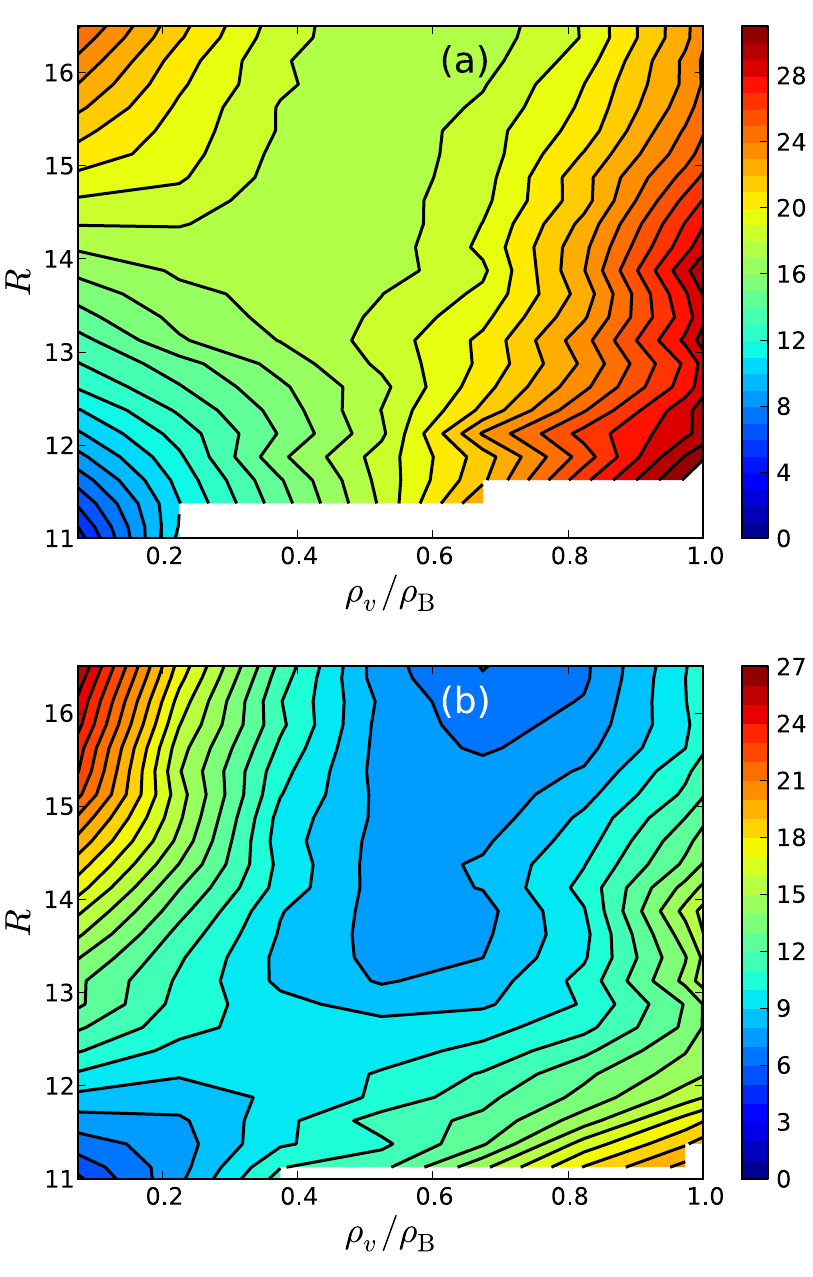} 
\caption{Free energy as a function of C60-C60 distance, $R$, and
density of water in the observation volume $v$ with respect to that in
the bulk, $\rho_v/\rhob$, for the association of two hydrophobic fullerene particles
in (a) GT and (b) GTRC water models. Contour lines are spaced in increments of $\kT$.}
\label{fig:2dpmf}
\end{figure}

The free energy surface shown in the top panel of Figure~\ref{fig:2dpmf} indicates that
hydrophobic collapse in GT water (or SPC/E water) is indeed driven by the barrier-less formation of a vapor
tube~\cite{LuzarVaporTube, WillardChandlerCGSolutes}
at  a C60-C60 distance between 14 and 15~\AA.
Hydrophobic collapse in GTRC water, on the other hand, does not follow this
mechanism because capillary evaporation in the inter-fullerene region has been suppressed
by the removal of LJ attractions in the solvent. This is consistent with the lack of drying at the interface of a single repulsive solute, as evidenced by
the nonuniform densities shown in the inset of Figure~\ref{fig:c60pmf}a and would be anticipated
from the results presented in Section IV.

Instead, the free energy minimum in GTRC water
(for a specific value of $R$) remains at liquid-like densities
as the C60-C60 distance is decreased, until the water molecules
cannot physically remain between the fullerene particles due to repulsive core overlap near
$R\approx12$~\AA. Only at this point are the solvation shell water molecules in the inter-fullerene region expelled.
This expulsion of water molecules in the observation volume causes the large free energy barrier
observed at the same inter-fullerene distance in the one-dimensional $W_{\rm solv}(R)$ for GTRC water shown
in Figure~\ref{fig:c60pmf}a. 

%@@@@@@@@@@@@@@@@@@@@@@@@@@@@@@@@@@@@@@@@@@@@@@
%@@@@@@@@@@@@@@@@@@@@@@@@@@@@@@@@@@@@@@@@@@@@@@

Instead of artificially suppressing capillary evaporation between large hydrophobes
by removal of solvent LJ attractions as in GTRC water, we can directly counteract the unbalanced LJ interfacial forces
leading to evaporation in the GT or full water models
by making the solutes sufficiently hydrophilic. LMF theory would predict very similar behavior for these two systems. 
This is accomplished
by using the full $U_{\rm FF}$ and $U_{\rm wF}$ potentials to describe
fullerene-fullerene and water-fullerene interactions, respectively.
Inclusion of the water-C60 attractive interactions leads to
an almost perfect cancellation of these unbalanced
forces, as evidenced by the
good agreement of the SPC/E and GT nonuniform
densities with that of the GTRC model, shown in the inset of Figure~\ref{fig:c60pmf}b.

These strong solute-water attractions, arising from the high surface density of carbon atoms,
render the C60 molecule hydrophilic, and the associated solvent-induced PMFs are repulsive for
all distances.  This indicates that water opposes the association of two such particles,
in accord with previous results~\cite{HphilC60PMF}.
Because of the effective hydrophilicity of the particles,
capillary evaporation between the particles
does not occur, and $W_{\rm solv}(R)$ is the same for all three models for $R\ge12$~\AA.
At smaller separations water is forcibly expelled from the inter-fullerene region due to overlap
with the repulsive cores of the solutes and then differences arise due to the differing pressure of the systems.

\begin{figure}
\centering
\includegraphics{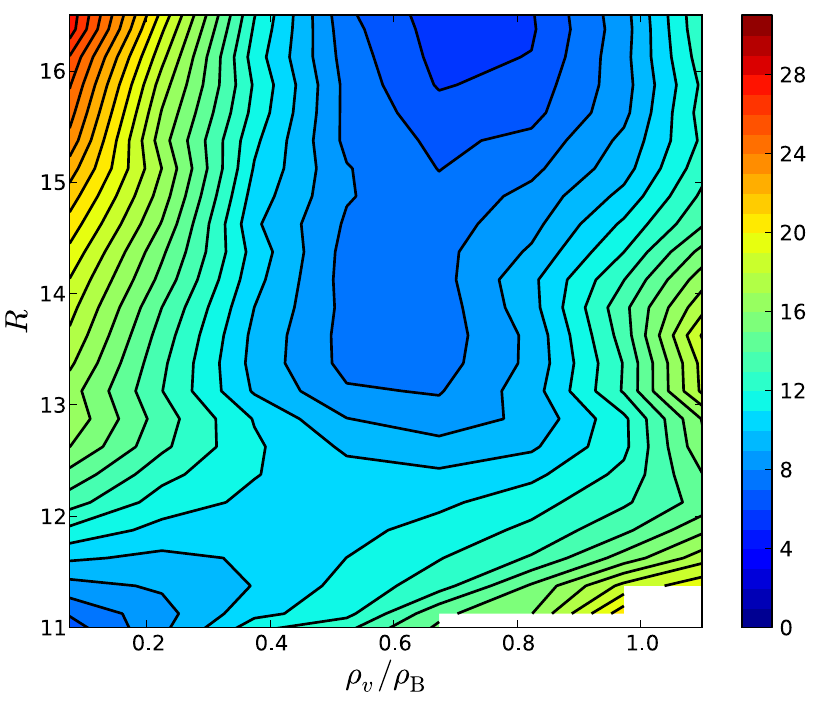} 
\caption{Free energy as a function of C60-C60 distance, $R$, and density
density of water in the observation volume $v$ with respect to that in
the bulk, $\rho_v/\rhob$, for the association of two hydrophilic fullerene particles in
GT water. Contour lines are spaced in increments of $\kT$.}
\label{fig:2dhphilpmf}
\end{figure}

The two-dimensional PMF $W(R,\rho_v)$ was also calculated
for the case of hydrophilic fullerene particles in GT water, and is shown
in Figure~\ref{fig:2dhphilpmf}. This PMF is qualitatively very similar to that shown for hydrophobic collapse
in GTRC water in Figure~\ref{fig:2dpmf}b as expected.  As $R$ is decreased, the free energy minimum as a function
of $\rho_v$ remains in regions of liquid-like densities. It is not until very small $R$, less than
12~\AA, that $W(R,\rho_v)$ develops a minimum at low $\rho_v$, indicating
a global free energy minimum at the contact state. 
In fact, the solvent induced PMF $W_{\rm solv}(R)$ between hydrophobic solutes in GTRC water is nearly
identical to the PMFs obtained between hydrophilic solutes in all models until water is expelled
from the inter-fullerene region, $R<12$~\AA, as illustrated by the curve labeled `GTRC-Hphob' 
in Figure~\ref{fig:c60pmf}b.
In contrast to what is found
for the association of large hydrophobic particles, the solvent opposes association
and the contact state is stabilized by the large solute-solute attractions between hydrophilic fullerenes.

%@@@@@@@@@@@@@@@@@@@@@@@@@@@@@@@@@@@@@@@@@@@@@@
%@@@@@@@@@@@@@@@@@@@@@@@@@@@@@@@@@@@@@@@@@@@@@@

\section{Conclusions}

We have used short ranged variants of the 
SPC/E water model~\cite{JStatPhys} in conjunction
with LMF theory to examine the crossover in the behavior of hydrophobic 
hydration with increasing solute size. 
While small scale solvation is determined exclusively by the 
local structure of water, \ie \ the hydrogen bond network, 
long ranged interactions are important for the accurate 
description of the hydration of large apolar solutes.
Dispersion interactions lead to the phenomena of drying at
extended hydrophobic interfaces,
while long ranged dipolar interactions are essential for the 
description of the orientational ordering of water
in the vicinity of a large solute, as well as for interfacial 
electrostatic properties. 

The truncated GT and GTRC water models also provide
insight into hydrophobic interactions between solutes in the small and large length scale regimes. 
The local structure of water, dictated by the hydrogen bond network, is found to govern
the association of two small scale solutes, a concept which has
been successfully exploited to provide a theoretical framework for describing
hydrophobic hydration and association at small length scales~\cite{HummerInfoTheory}.
Moreover, previous work has shown
that coarse-grained models, whereby water molecules interact via a
single spherically symmetric pairwise potential, can reproduce the thermodynamics
of association of two methanes~\cite{SSWater,SSWaterMethane}.
From the results presented here, it is not surprising
that such coarse-grained models can capture features of small scale hydrophobicity,
since these models also describe the bulk structure of water with near quantitative accuracy.

The association of two large scale hydrophobes involves the formation
of an inter-solute vapor tube, and the unbalanced forces arising from water-water
LJ attractions are found to be of the utmost importance for this mechanism
of hydrophobic association. In this regime the coarse-grained water models will fail completely.
Cancellation of the effects of interfacial unbalanced
forces, either by explicit removal of solvent-solvent LJ attractions (as in GTRC water)
or by addition of large solute-water attractions that counterbalance these 
forces, suppresses capillary evaporation between two large solutes.
As a result the solute surface is wet by the aqueous solvent,
and free energy barriers to the association
of two large hydrophilic solutes exist. In all these cases comparison of results in the full model with those from the short-ranged GT and GTRC water models provides a simple and physically suggestive way to disentangle the effects of longer ranged dispersive and Coulomb interactions from properties of the local hydrogen bond network. 

 \begin{acknowledgments}
This work was supported by the National Science Foundation
(grant CHE0848574). We are grateful to Shekhar Garde, Hari Acharya,
David Chandler, and Jocelyn Rodgers for many helpful discussions.
\end{acknowledgments}

%\begin{tocentry}

%Some journals require a graphical entry for the Table of Contents.
%This should be laid out ``print ready'' so that the sizing of the
%Inside the \texttt{tocentry} environment, the font used is Helvetica
%8\,pt, as required by \emph{Journal of the American Chemical
%The surrounding frame is 9\,cm by 3.5\,cm, which is the maximum
%permitted for  \emph{Journal of the American Chemical Society}
%graphical table of content entries. The box will not resize if the
%content is too big: instead it will overflow the edge of the box.
%
%This box and the associated title will always be printed on a
%separate page at the end of the document.
%\newpage

%@@@@@@@@@@@@@@@@@@@@@@@@@@@@@@@@@@@@@@@@@@@@@@
%\includegraphics[width=0.5\textwidth]{TOC-1.pdf} 
%\\
%{Table of Contents Graphic}
%@@@@@@@@@@@@@@@@@@@@@@@@@@@@@@@@@@@@@@@@@@@@@@

%\end{tocentry}

\end{document}